\documentstyle[preprint,aps]{revtex}

\begin{document}

\begin{center} The Love-Hate Relationship Between the
Shell Model and Cluster Models \medskip \\ 
Larry Zamick and Shadow J.Q. Robinson

\noindent  Department of Physics and Astronomy,\\
Rutgers University, Piscataway, \\
New Jersey  08854-8019\\
Proceedings of the International Conference ``Clustering 
Phenonema in Nuclear Physics'',\\
St. Petersburg, Russia June 14-17
\end{center}

\bigskip
\begin{abstract}
We adopt a personal approach here reviewing several calculations 
over the years in which we have experienced confrontations between cluster 
models and the shell model.  In previous cluster conferences we have noted 
that cluster models go hand in hand with Skyrme Hartee-Fock calculations in 
describing states which cannot easily, if at all, be handled by the shell model.
These are the highly deformed (many particle - many hole) intruder states, 
linear chain states e.t.c. In the present work we will consider several 
topics; the quadrupole moment of $^{6}$Li, the non-existence of low 
lying intruders in $^{8}$Be, and then jumping to the $f_{7/2}$ shell, 
we discuss the two-faceted nature of the nuclei - sometimes displaying 
shell model properties, other times cluster properties.
\end{abstract}

\section{The quadrupole moment of the J = 1$^{+}$ state in $^{6}$Li}
Whereas the quadrupole moment of the deuteron is positive(Q = +2.74mb), that of 
the J=1$^{+}$ state of $^{6}$Li is negative, Q=-0.818(17)mb.  
The magnetic moment of the deuteron is $\mu=0.85741 $ nm while that of $^{6}$Li is 0.822 nm.

There appears to be a big discrepancy between cluster model calculations 
and the shell model calculations. In nearly all cluster model calculations 
Q comes out \underline{positive}.  However in many shell model calculations 
Q comes out \underline{negative}, sometimes too negative.  This is an 
important problem that deserves further attention.  See for example arguments in the literature between the cluster group[1] 
and the shell model group[2].  See also the recent compendium of A=6 by D.R. Tilley et. al.[3].

For example in a modern shell model approach by Forest et. al. [4]
 gets about -8 mb for Q, a factor of 10 
too large but of the correct sign.  On the other hand in a 
dynamical microscopic three cluster description of $^{6}$Li 
where the clusters are $\alpha$, n, and p the result is Q = 2.56 mb.[1]

In shell model calculations that we performed [5] we started with 2 valence particles 
in the 0p shell (0$\hbar \omega$).  Then we allowed up to 2 $\hbar \omega$ and then up 
to 4 $\hbar \omega$ excitations.  In the 0 $\hbar \omega$ space if you do not have a 
tensor interaction Q comes out positive.  With a 'realistic' tensor interaction 
Q comes out negative but too negative Q=-3.5mb.  However with a former 
student Zheng, who at Arizona also developed the no core approximation 
with Barrett et. al.[6], we showed that when higher shell admixtures were admitted Q became 
smaller in magnitude and closer to experiment as shown in the following table.[5]

The results are shown in the following table
\begin{center}
\begin{tabular}{ccc}
SPACE &  Q(mb) & $\mu$(mm) \\
\tableline
\tableline
0 $\hbar \omega$ &  -3.60 & 0.866\\
2 $\hbar \omega$ &  -2.51 & 0.848\\
4 $\hbar \omega$ &  -0.085& 0.846\\
Experiment       &  -0.82 & 0.822\\
\end{tabular}
\end{center}

Note that the shell model calculations cannot get the 
magnetic moment low enough.  With up to 4
$\hbar \omega$ admixtures we actually overshoot and get a 
quadrupole moment that is too small but still negative.  Some
cluster models appear to explain the low magnetic moment.  

An excellent discussion of many shell model calcultion of Q and $\mu$ has 
been given by Karataglidis et. al [7].  The value of Q that they obtain with what they call the ``Zheng'' interaction [8]
in the up to 0,2,4 and 6 $\hbar\omega$ spaces are -2.64,-2.08,-0.12, and 0.17 mb respectively.  Thus they get Q to become positive at the 
6 $\hbar\omega$ level.  But then they quote Zheng et. al [8] as getting a value of -0.67mb in the same 6 $\hbar\omega$ space.  It is 
not clear why the two calculations give different answers.  The changes in 
$\mu$ in ref [7] are more moderate 0.869,0.848, 0.845 and 0.840 nm in the 
up to 0,2,4 and 6 $\hbar\omega$ spaces.

Looking at all the calculations by all groups (including our own), the situation is certainly 
confusing, and the problem deserves further attention.  This is 
certainly a \underline{basic} problem, the deuteron embedded in the 
nuclear medium.  This problem has wider implications whether or not 
there is T=0 pairing can depend on how higher order configurations 
affect the tensor interaction in the valence space.

\section{Absence of Low Lying Intruders in $^{8}$Be and the $\alpha$ particle model}

The 0$^{+}$ bandhead for low lying intruders in $^{16}$0, $^{12}$C, 
and $^{10}$Be are at 6.05 MeV, 7.65 MeV and 6.18 MeV respectively.  
In $^{16}$0 and  $^{12}$C, these are predominantly 4 particle 4 hole 
excitations.  In $^{12}$C, we identify the intrinsic state as a 
linear chain.  In the 7th edition of the 'Tables of Isotopes' 
\underline{possible} intruders in $^{8}$ Be were indicated, a 
J=0$^{+}$ state at 6 MeV and J=2$^{+}$ state at 9 MeV.

In shell model calculations allowing 2 particle 2 hole excitations 
we were able with a quadrupole-quadrupole force to get a J=0$^{+}$ 
state at 9.7 MeV in $^{10}$Be, too high but in the right ballpark.[7,8]  But we could 
not get low lying intruders in $^{8}$Be below 30 Mev.[7,8]  We used a 
deformed oscillator model to show why one gets intruders in 
$^{12}$C and $^{10}$Be but not $^{8}$ Be.

But perhaps the simplest explanation as suggested to us by E. Vogt is given by the $\alpha$ 
particle model.  In $^{12}$C we can rearrange the $\alpha$ 
particles from a triangle to a linear chain.  In $^{8}$Be we 
have only 2 $\alpha$ particles.  One can get a rotational band 
by having the 2 $\alpha$'s rotate around each other but that is all. 

The mere existence of these intruder states is of astrophysical 
importance.  In the beta decay $^{8}$B$ \rightarrow ^{8}$Be + e$^{+}+ \nu$ 
one goes from a J=1$^{+}$ T=1 to J=2$^{+}$ states. This is the famous 'Ray Davis' neutrino. If there 
\underline{were} a 2$^{+}$ state at 9 MeV then there would be more 
high energy $\alpha$'s than there would be if the decay were 
to the $2_{1}^{+}$ state at 3.04 MeV.[9]  The alpha spectrum from the decay of $^{8}$Be seems to show more high energy alphas,
but we would say that they are not due to low lying intruders. 

\section{Clustering and shell model in the $f_{7/2}$ region}

In a previous cluster conference in Santorini  (1993) a spectrum of $^{44}$Ti was
shown in an $\alpha$ cluster model.[10]  The spectrum looked reasonable except
that there was a wide gap between the $10^{+}$ and $12^{+}$ states.  However,
these states are sufficiently close together that the $12^{+}$ state is isomeric.
In a single j shell basis (j=f$_{7/2}$) $^{52}$Fe is the 4 hole system and it should
have an identical spectrum to that of $^{44}$Ti provided the same interaction
is used.  However in $^{52}$Fe the $12^{+}$ lies below the $10^{+}$ state.
It is extremely isomeric and has a lifetime of 12 minutes.

We have studied this and other topics by calculating the spectrum of $^{44}$Ti ($^{52}$Fe)
with a variety of interactions designated as Model X. (See Tables I,II)

Model I:  Use the spectrum of $^{42}$Sc as input (particle-particle) Identify $<(j^{2})^{J}V(j^{2})^{J}>=E(J)$
experimental. For isospin T=0 J can be 1,3,5 and 7 while for T=1 J is even 0,2,4, and 6.

Model II:  Use the spectrum of $^{54}$Co as input (hole - hole).  If there were no
configuration mixing these two spectra would be identical.  However, there
are some differences eg the $7^{+}$ state is much lower in $^{54}$Co than in $^{42}$Sc.

Model III:  Now we play games.  We want to find out how important are the T=0 matrix
elements for the structure of the nuclei. (e.g. Is T=0 pairing important?)  Noticing that in $^{42}$Sc
the J=2,3 and 5 states are nearly degenerate in this model we set all
the T=0 matrix elements to be the same and to all equal $E(2^{+})=1.5863$ MeV.

In model III we then have $V^{T=1}= V(^{42}Sc)^{T=1}$ J =0,2,4,6 and $V^{T=0}=$ constant$=E(2^{+})$ J=1,3,5,7.
We can then write $V^{T=0}$=constant $(1/4 - t_{1} \cdot t_{2})$.
We can then write $V^{T=0}=c(1/4- t_{1}\cdot t_{2})$ where c is a constant.  Hence 
$\sum_{i<j}V_{ij}^{T=0}=c/8(n(n-1)+6)-c/2T(T+1)$.  This means that the spectrum of states of a given isospin e.g. T=0
in $^{44}$Ti($^{52}$Fe) is independent of what the constant is, it
might as well be \underline{zero}.  Of course the relative
splitting of T=1 and T=0 states will be affected.  Model III
will be the standard from which we derive Model IV.

Model IV:  Relative to the degenerate case above, we now move the $J=1^{+}$
state down in energy to 0.5863 MeV.  Our motivation is based on
numerous discussions about the importance of T=0 S=1 ``pairing''
in nuclei.  We hope to simulate the T=0 pairing by this lowering.

Model V:  Relative to Model III we bring the J=1$^{+}$ and J=7$^{+}$ states down to an 
energy of 0.5863 MeV but keep the J=3$^{+}$ and 5$^{+}$ at E=E(2$^{+}$)=1.5863 MeV.  
This spectrum is very close to that of $^{42}$Sc.

\section{Discussion of Results}

Let us first compare Model III (all T=0 matrix elements are degenerate) with Model I (spectra of $^{42}$Sc).
As already mentioned, making T=0 matrix elements degenerate is equivalent
to making them zero as far as T=0 states are concerned.

The main difference is that the states with J=6,4,7, and 8 come down in energy
as does J=9$^{+}$.  Also the 12-10 gap is a bit greater than for the $^{42}$Sc spectra
case, reminiscent of the $\alpha$ particle model.  The J=9$^{+}$ state is below
the 10$^{+}$ and 12$^{+}$ in the degenerate case.  Clearly it is the high energy side of
the spectrum which is most sensitive to the change from experimental
spectrum to the ``T=0 degenerate'' case.

Despite the changes, we can say that the T=1 two body matrix elements give
the dominant structure of the spectrum whilst the T=0 matrix elements
provide the fine tuning.

We now compare Model IV with Model III. The only difference is that we break the T=0
degeneracy by lowering the J=$1^{+}$ state from 1.5863 MeV to 0.5863 Mev.
We hope that this simulates to some extent T=0 S=1 neutron-proton pairing.
The change from degenerate case is not that large.  There is a tendency to
go towards the spectrum of $^{42}$ Sc.  The J=3,5,7, and 8  states are raised
somewhat in energy.  However it is hard to find a clear signature of this S=1 pairing.

Not shown is Model V where we bring down both the J=1$^{+}$ and 7$^{+}$ 
states to 0.5863 MeV
keeping J=3 and 5 at E(2) = 1.5863 MeV.  This input spectrum is close to that of
$^{42}$ Sc so that it is not surprising that the $^{44}$Ti spectrum is likewise close.

We lastly consider the results using the spectrum of $^{54}$Co.  This was done
some time ago by Geesaman [11,12]  Note that there are significant changes, all at
the high energy high angular momentum part of the spectrum.
Relative to the $^{42}$Sc case the 10$^{+}$ and 12$^{+}$ states are down in energy with
the 12$^{+}$ below the 10$^{+}$ thus leading to a long lifetime for the 12$^{+}$ state.
Note that the 9$^{+}$ state is now at a much high energy than the 10$^{+}$ or 12$^{+}$.  Recently
the 10$^{+}$ state in $^{52}$Fe, which lies above the 12$^{+}$ has been found in $^{52}$Fe 
by Ur et. al. [13].  It would also be of interest to find the 9$^{+}$ state.

\section{Many particle, Many Hole States in $^{40}$Ca }
This is a topic we discussed in previous cluster meetings so 
we will be brief.[14]  We just want to remind the reader that 
there are all sorts of many particle-many hole highly deformed 
states in $^{40}$Ca.  One cannot properly describe $^{40}$Ca 
in a cluster model consisting of $^{36}$Ar plus an alpha 
particle.  At the very least one has to start with $^{32}$S 
plus two alpha particles.

In a Skyrme Hartree-Fock calculations (SK III) we obtain 
a near degeneracy of the 4p-4h and 8p - 8h intrinsic state.
  The respective energies are 12.1 MeV and 11.4 MeV.  
The 8p-8h intrinsic state energy is lower than the 4p-4h.
By the time projection and pairing are included, the 4p-4h 
comes lower than the 8p-8h (6.85 MeV vs 8.02 MeV) in 
agreement with the order the J = 0 excitation energies of 
3.0 and 5.1 Mev.  Pairing will lower the states even more.  
We actually found many more deformed states of the form 
np-nh n=2,3,4,5,6,7 and 8.  The intrinsic states are 
nearly degenerate in energy - we called this a deformation 
condensate.  We also found for these states that the deformation 
parameter was approximately proportional to n i.e. the value of 
$\beta$ for 8p-8h is approximately twice the value of $\beta$ for 4p-4h.

In $^{80}$Zr one of the np-nh states becomes the ground state.
  This is the 12p-12h state which has a calculated value of 
$\beta=0.4$  A more superdeformed 16p-16h state with $\beta=0.6$ 
is calculated to be at an excitation energy of about 8 MeV.

\section{Two different Views of the $f_{7/2}$ region}

In March 2000 issue of the Physical Review C \underline{61} there 
are two papers side by side.  One is by our group [15] and one 
by H. Hasegawa and K. Koneko [16].  We both do calculations in 
the f$_{7/2}$ shell.  We emphasize shell model behavior whilst 
the other authors the $\alpha$ cluster behaviors, even though their 
model space is limited to f$_{7/2}$.

The other authors point out that we can get an excellent approximation 
to the ground states of n$_{p}$ = n$_{n}$=2m nuclei (n$_{p}$ is the 
number of protons e.t.c.). 
\begin{equation}
|(f_{7/2})^{4m}I=T=0> = \frac{1}{\sqrt{N_{0}}} (\alpha_{0}^{\dagger})^{m}|A_{0}>
\end{equation}
where $(\alpha_{0}^{\dagger})$ creates a 4 nucleon cluster.
\begin{equation}
\alpha^{\dagger}_{0}=\sum_{J,\tau} (J\tau,J\tau:I=T=0)(A^{\dagger}_{J\tau}A^{\dagger}_{J\tau})_{I=0 T=0}
\end{equation}
For $^{48}$Cr this approximation give -32.04 MeV for the ground 
state energy where as the exact value is 32.70 MeV.

We on the other hand have emphasized the shell model aspects.[15]  
In the previously mentioned paper, we find an approximation for 
the excitation energies of single and double analog states in 
the f$_{7/2}$ region and in a an earlier paper ``Fermionic 
Symmetries: Extension of the two to one relationship between 
spectra of even even and neighboring odd mass nuclei''[17] we noted 
two things.\\
A. There is often a two to one relation between spectra of even-even 
and even odd nuclei, and in some cases the single j shell model predicts this.\\
B. Excitation energies of analog state are approximately the same if the neutron excess (or equivalently the ground state isospin) is the same.

The above results can be \underline{parametrized} by the following formulae\\
SINGLE ANALOG EXCITATION (SA)
\begin{equation}
E(SA)=b(T+X)
\end{equation}
DOUBLE ANALOG EXCITATION (DA)
\begin{equation}
E(DA)=2b(T+X+1/2)
\end{equation}
This formula will give a two to one ratio for E(DA)/E(SA) for ($^{44}$Ti,$^{43}$Ti),
($^{51}$Cr,$^{50}$Cr), ($^{47}$Sc,$^{48}$Ti) etc.

The experimental SA and DA are shown in the following table.

\begin{tabular}{ll}
T=0 	& 	$^{44}$Ti (9.340), $^{48}$Cr (8.75), $^{52}$Fe (8.559) \\
T=1/2 	& 	$^{43}$Sc (4.274)$^{a}$, $^{43}$Ti (4.338)$^{a}$,
		$^{45}$Ti (4.176), $^{49}$Cr(4.49), \\
  	&	$^{51}$Mn (4.451), $^{53}$Co (4.390), $^{53}$Fe (4.250) \\
T=1 	& 	$^{46}$Ti (14.153), $^{50}$Cr (13.222) \\
T=3/2 	& 	$^{45}$Sc (6.752)$^{a}$, $^{47}$Ti (7.187),
		$^{51}$Cr (6.611) \\
T=2 	& 	$^{48}$Ti (17.379) \\
T=5/2 	& 	$^{47}$Sc (8.487)$^{a}$, $^{49}$Ti (8.724)\\
\end{tabular}

$a$ - obtained from binding energy data.

In Table II we compare the \underline{theoretical} single j 
shell calculations with the linear formula.  We take 
b = 2.32 MeV X=1.30.  Note that in the SU(4) limit X=2.5.  
The fact that SU(3) gives the linear formula is not sufficient 
for it to be the correct theory.  For a simple monopole-monopole 
interaction a+bt(1)t(2) X=1.  

Some of the two to one ratio's hold rigoursly in the single j 
shell model.  This holds for 3 particle and 4 particle systems 
or 3 holes and 4 holes.  eg ($^{43}$Ti,$^{44}$Ti),
($^{43}$Sc,$^{44}$Ti),($^{53}$Fe,$^{52}$Fe).   Here not only 
single or double analog but all the J=j states in the odd 
spectrum are at half the energy of the J=0+ states in the even system.

Some of the relations hold approximately in the single j shell model 
eg for ($^{45}$Sc,$^{46}$Ti) and for the cross conjugate pair 
($^{51}$Cr,$^{50}$Cr) we would get a 2 to 1 ratio if seniority four states 
could be neglected and one only had v=0 and v=2.

Miraculously the 2 to 1 ratio holds remarkably well \underline{experimentally} 
for ($^{51}$Cr,$^{50}$Cr) - the values are 6.511 and 13.022 MeV respectively, 
this despite the fact that in the single j shell it should only hold approximately.  
Ironically the simplest system for which 2 to 1 should hold exactly does 
not work so well.  That is to say for ($^{43}$Ti,$^{44}$Ti) the values are 
4.338 and 9.340 MeV.  When configuration mixing is included, agreement with 
the deviation is explained.  This might be an example of a 4 particle clustering.
  For the hole system ($^{53}$Fe, $^{52}$Fe) on the other hand 2 to 1 works much better.

The fact that there is in general a close relation between even-even 
and even-odd puts to question whether in those many cases there is 
any $\alpha$ particle clustering.

The single j shell calculation does not predict exact equality for the S.A. excitation energies in 
$^{43}$Sc and $^{45}$Ti.  The relative values are very close however 4.142 and 4.112 MeV respectively.  This is fascinating.  
We take $^{43}$Sc, jam a deuteron into it to form $^{45}$Ti and it seems hardly to make any difference for S.A. excitations.[15]

\section{Two views of cross conjugate relations}

In the single j shell model, the spectra of cross conjugate nuclei 
should be identical.[18]  (for j$^{n}$ states.). A cross conjugate nucleus 
is one obtained by changing protons into neutron holes and neutrons 
into proton holes.  The cross conjugate of $^{46}$Ti is $^{50}$Cr.  
Let us compare the spectra
 
\begin{tabular}{ccccccc}
J & \hspace{2cm} &$^{46}$Ti & \hspace{2cm} &$^{50}$Cr& \hspace{2cm} & Ratio  \\
0 & \hspace{2cm} &0.000     & \hspace{2cm} &0.000    & \hspace{2cm} &        \\
2 & \hspace{2cm} &0.889     & \hspace{2cm} &0.787    & \hspace{2cm} & 0.8853 \\
4 & \hspace{2cm} &2.010     & \hspace{2cm} &1.884    & \hspace{2cm} & 0.9373 \\
6 & \hspace{2cm} &3.297     & \hspace{2cm} &3.164    & \hspace{2cm} & 0.9597 \\
8 & \hspace{2cm} &4.896     & \hspace{2cm} &4.740    & \hspace{2cm} & 0.9681 \\
\end{tabular}
\\

The fits are very good.  The $^{50}$Cr excitations are slightly smaller - it could be a universal A dependence.  Where 
does the remarkable agreement leave room for $\alpha$ clustering?

However we can look for other things besides the spectra.  In a recent 
experiment theory collaboration where the leading experimentalists 
were N. Koller and H.A. Speidel [19] good agreement was obtained 
for g(2$^{+}$) in $^{50}$Cr but bad agreement for $^{46}$Ti.  
The shell model predicts a high value of g(2$^{+}$) and g(4$^{+}$) for $^{50}$Cr but 
low values for $^{46}$Ti $~$ 0.25 mm.  The high values are confirmed for $^{50}$Cr 
but for $^{46}$Ti the measured g factors are closer to 0.5 which suggest the 
rotational value g$_{R}$=Z/A.  These results suggest that there 
must be considerable clustering in $^{46}$Ti that is not present in $^{50}$Cr.  In
general the shell model appears to work better in the upper half of the ``f$_{7/2}$ shell''
than in the lower half.  There appears to be much more going on in the lower half and probably
this is due to intruder/cluster mixing in with the basic shell model states.

\section{Closing Remarks}

We have provided several examples where Cluster Models and the Shell Model confront 
each other usually to the mutual benefit
of both models even though in the short term there might be some arguments.  
The two models give the opposite sign for the quadrupole moment 
of $^{6}$Li, and this has to be resolved.  The cluster model provides insight 
into some results of detailed shell model calculations
e.g. why there are no low lying intruders in $^{8}$Be.  The low lying intruder 
states e.g. 4p-4h and 8p-8h in $^{40}$Ca are essentially impossible to calculate 
in the shell model.  However here cluster models and Skyrme-Hartree Fock go 
together in describing such states.  In the f$_{7/2}$ region we raise the 
question (without fully answering it) of how to distinguish symmetry energy 
from clustering energy.  Finally we point out the issue of hidden clustering. 

This work was supported in part by the U.S. Department of Energy DE-FG02-95ER-40940

\setlength{\unitlength}{3947sp}%
\begingroup\makeatletter\ifx\SetFigFont\undefined
\def\x#1#2#3#4#5#6#7\relax{\def\x{#1#2#3#4#5#6}}%
\expandafter\x\fmtname xxxxxx\relax \def\y{splain}%
\ifx\x\y   
\gdef\SetFigFont#1#2#3{%
  \ifnum #1<17\tiny\else \ifnum #1<20\small\else
  \ifnum #1<24\normalsize\else \ifnum #1<29\large\else
  \ifnum #1<34\Large\else \ifnum #1<41\LARGE\else
     \huge\fi\fi\fi\fi\fi\fi
  \csname #3\endcsname}%
\else
\gdef\SetFigFont#1#2#3{\begingroup
  \count@#1\relax \ifnum 25<\count@\count@25\fi
  \def\x{\endgroup\@setsize\SetFigFont{#2pt}}%
  \expandafter\x
    \csname \romannumeral\the\count@ pt\expandafter\endcsname
    \csname @\romannumeral\the\count@ pt\endcsname
  \csname #3\endcsname}%
\fi
\fi\endgroup
\begin{picture}(6402,5589)(439,-7711)
\thinlines
\put(451,-7261){\line( 1, 0){900}}
\put(451,-6811){\line( 1, 0){900}}
\put(451,-6046){\line( 1, 0){900}}
\put(451,-5416){\line( 1, 0){900}}
\put(451,-4696){\line( 1, 0){900}}
\put(451,-4651){\line( 1, 0){900}}
\put(451,-4561){\line( 1, 0){900}}
\put(451,-4471){\line( 1, 0){900}}
\put(451,-3976){\line( 1, 0){900}}
\put(451,-3796){\line( 1, 0){900}}
\put(451,-3661){\line( 1, 0){900}}
\put(2251,-7261){\line( 1, 0){900}}
\put(2251,-6811){\line( 1, 0){900}}
\put(2251,-6091){\line( 1, 0){900}}
\put(2251,-5416){\line( 1, 0){900}}
\put(2251,-4696){\line( 1, 0){900}}
\put(2251,-4561){\line( 1, 0){900}}
\put(2251,-4381){\line( 1, 0){900}}
\put(2251,-4336){\line( 1, 0){900}}
\put(2251,-4291){\line( 1, 0){900}}
\put(2251,-4246){\line( 1, 0){900}}
\put(2251,-3616){\line( 1, 0){900}}
\put(4051,-7261){\line( 1, 0){900}}
\put(5851,-7261){\line( 1, 0){900}}
\put(4051,-6676){\line( 1, 0){900}}
\put(4051,-6046){\line( 1, 0){900}}
\put(4051,-5686){\line( 1, 0){900}}
\put(4051,-5191){\line( 1, 0){900}}
\put(4051,-5011){\line( 1, 0){900}}
\put(4051,-4876){\line( 1, 0){900}}
\put(4051,-4741){\line( 1, 0){900}}
\put(4051,-4021){\line( 1, 0){900}}
\put(4051,-3751){\line( 1, 0){900}}
\put(5851,-6676){\line( 1, 0){900}}
\put(5851,-6001){\line( 1, 0){900}}
\put(5851,-5596){\line( 1, 0){900}}
\put(5851,-5011){\line( 1, 0){900}}
\put(5851,-4966){\line( 1, 0){900}}
\put(5851,-4696){\line( 1, 0){900}}
\put(5851,-4561){\line( 1, 0){900}}
\put(5851,-3796){\line( 1, 0){900}}
\put(5851,-3526){\line( 1, 0){900}}
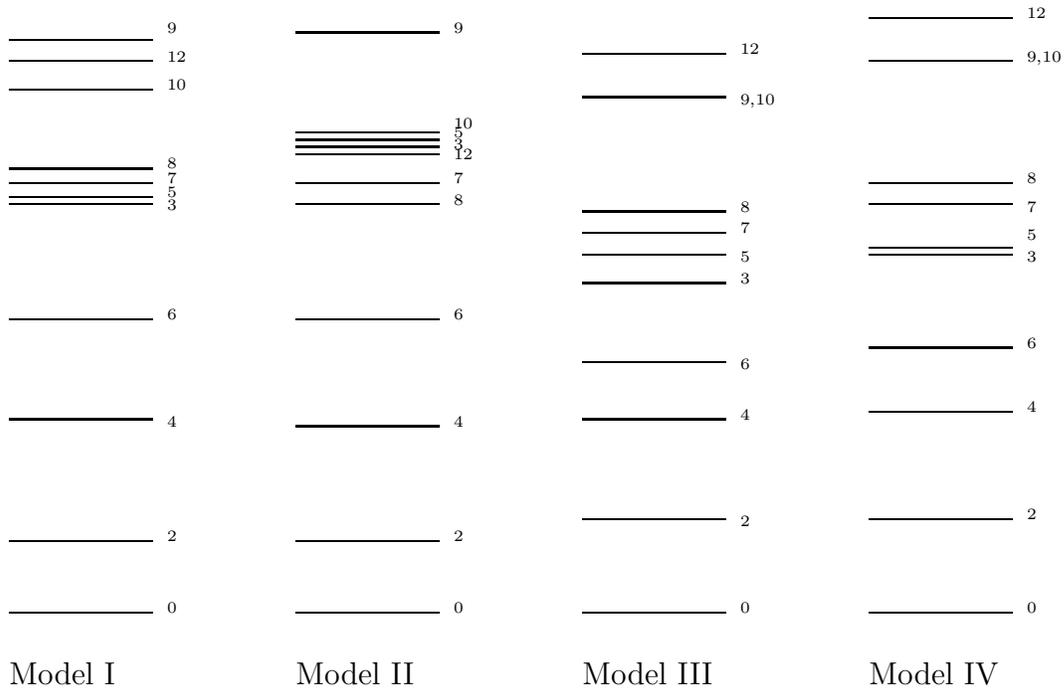
\put(2746,-3266){\makebox(0,0)[lb]{\smash{\SetFigFont{12}{14.4}{rm}Ti 44 spectra for Models I,II,III, and IV }}}
\put(451,-7711){\makebox(0,0)[lb]{\smash{\SetFigFont{12}{14.4}{rm}Model I}}}
\put(2251,-7711){\makebox(0,0)[lb]{\smash{\SetFigFont{12}{14.4}{rm}Model II}}}
\put(4051,-7711){\makebox(0,0)[lb]{\smash{\SetFigFont{12}{14.4}{rm}Model III}}}
\put(5851,-7711){\makebox(0,0)[lb]{\smash{\SetFigFont{12}{14.4}{rm}Model IV}}}
\put(6841,-7261){\makebox(0,0)[lb]{\smash{\SetFigFont{6}{12.0}{rm}0}}}
\put(6841,-6676){\makebox(0,0)[lb]{\smash{\SetFigFont{6}{12.0}{rm}2}}}
\put(6841,-6001){\makebox(0,0)[lb]{\smash{\SetFigFont{6}{12.0}{rm}4}}}
\put(6841,-5596){\makebox(0,0)[lb]{\smash{\SetFigFont{6}{12.0}{rm}6}}}
\put(6841,-5056){\makebox(0,0)[lb]{\smash{\SetFigFont{6}{12.0}{rm}3}}}
\put(6841,-4921){\makebox(0,0)[lb]{\smash{\SetFigFont{6}{12.0}{rm}5}}}
\put(6841,-4741){\makebox(0,0)[lb]{\smash{\SetFigFont{6}{12.0}{rm}7}}}
\put(6841,-4561){\makebox(0,0)[lb]{\smash{\SetFigFont{6}{12.0}{rm}8}}}
\put(6841,-3796){\makebox(0,0)[lb]{\smash{\SetFigFont{6}{12.0}{rm}9,10}}}
\put(6841,-3526){\makebox(0,0)[lb]{\smash{\SetFigFont{6}{12.0}{rm}12}}}
\put(5041,-7261){\makebox(0,0)[lb]{\smash{\SetFigFont{6}{12.0}{rm}0}}}
\put(5041,-6721){\makebox(0,0)[lb]{\smash{\SetFigFont{6}{12.0}{rm}2}}}
\put(5041,-6046){\makebox(0,0)[lb]{\smash{\SetFigFont{6}{12.0}{rm}4}}}
\put(5041,-5731){\makebox(0,0)[lb]{\smash{\SetFigFont{6}{12.0}{rm}6}}}
\put(5041,-5191){\makebox(0,0)[lb]{\smash{\SetFigFont{6}{12.0}{rm}3}}}
\put(5041,-5056){\makebox(0,0)[lb]{\smash{\SetFigFont{6}{12.0}{rm}5}}}
\put(5041,-4876){\makebox(0,0)[lb]{\smash{\SetFigFont{6}{12.0}{rm}7}}}
\put(5041,-4741){\makebox(0,0)[lb]{\smash{\SetFigFont{6}{12.0}{rm}8}}}
\put(5041,-4066){\makebox(0,0)[lb]{\smash{\SetFigFont{6}{12.0}{rm}9,10}}}
\put(5041,-3751){\makebox(0,0)[lb]{\smash{\SetFigFont{6}{12.0}{rm}12}}}
\put(3241,-7261){\makebox(0,0)[lb]{\smash{\SetFigFont{6}{12.0}{rm}0}}}
\put(3241,-6811){\makebox(0,0)[lb]{\smash{\SetFigFont{6}{12.0}{rm}2}}}
\put(3241,-6091){\makebox(0,0)[lb]{\smash{\SetFigFont{6}{12.0}{rm}4}}}
\put(3241,-5416){\makebox(0,0)[lb]{\smash{\SetFigFont{6}{12.0}{rm}6}}}
\put(3241,-4696){\makebox(0,0)[lb]{\smash{\SetFigFont{6}{12.0}{rm}8}}}
\put(3241,-4561){\makebox(0,0)[lb]{\smash{\SetFigFont{6}{12.0}{rm}7}}}
\put(3241,-4410){\makebox(0,0)[lb]{\smash{\SetFigFont{6}{12.0}{rm}12}}}
\put(3241,-4351){\makebox(0,0)[lb]{\smash{\SetFigFont{6}{12.0}{rm}3}}}
\put(3241,-4277){\makebox(0,0)[lb]{\smash{\SetFigFont{6}{12.0}{rm}5}}}
\put(3241,-4221){\makebox(0,0)[lb]{\smash{\SetFigFont{6}{12.0}{rm}10}}}
\put(3241,-3616){\makebox(0,0)[lb]{\smash{\SetFigFont{6}{12.0}{rm}9}}}
\put(1441,-7261){\makebox(0,0)[lb]{\smash{\SetFigFont{6}{12.0}{rm}0}}}
\put(1441,-6811){\makebox(0,0)[lb]{\smash{\SetFigFont{6}{12.0}{rm}2}}}
\put(1441,-6091){\makebox(0,0)[lb]{\smash{\SetFigFont{6}{12.0}{rm}4}}}
\put(1441,-5416){\makebox(0,0)[lb]{\smash{\SetFigFont{6}{12.0}{rm}6}}}
\put(1441,-4731){\makebox(0,0)[lb]{\smash{\SetFigFont{6}{12.0}{rm}3}}}
\put(1441,-4651){\makebox(0,0)[lb]{\smash{\SetFigFont{6}{12.0}{rm}5}}}
\put(1441,-4561){\makebox(0,0)[lb]{\smash{\SetFigFont{6}{12.0}{rm}7}}}
\put(1441,-4471){\makebox(0,0)[lb]{\smash{\SetFigFont{6}{12.0}{rm}8}}}
\put(1441,-3976){\makebox(0,0)[lb]{\smash{\SetFigFont{6}{12.0}{rm}10}}}
\put(1441,-3796){\makebox(0,0)[lb]{\smash{\SetFigFont{6}{12.0}{rm}12}}}
\put(1441,-3616){\makebox(0,0)[lb]{\smash{\SetFigFont{6}{12.0}{rm}9}}}
\end{picture}

\begin{table}
\caption{Two particle Matrix Elements $<(j^{2})^{J}V(j^{2})^{J}>$}
\begin{tabular}{ccccccccc}
\tableline
Case  & J=0 T=1 & J=2 T=1 & J=4 T=1 & J=6 T=1 & J=1 T=0 & J=3 T=0 & J=5 T=0 & J=7 T=0 \\
\tableline
\tableline
Model I$^{a)}$		& 0.000 & 1.5863 & 2.8153 & 3.2420 & 0.6110 & 1.4904 & 1.5101 & 0.6163 \\
Model II$^{b)}$      	& 0.000 & 1.4465 & 2.6450 & 2.9000 & 0.9372 & 1.8224 & 2.1490 & 0.1990 \\
Model III$^{c)}$     	& 0.000 & 1.5863 & 2.8153 & 3.2420 & 1.5863 & 1.5863 & 1.5863 & 1.5863 \\
Model IV$^{d)}$       	& 0.000 & 1.5863 & 2.8153 & 3.2420 & 0.5863 & 1.5863 & 1.5863 & 1.5863 \\
\tableline
\tableline
\end{tabular}
\end{table}

\begin{table}
\caption{$^{44}$Ti ($^{52}$Fe) spectra for Model I,II,III and IV}
\begin{tabular}{cccc}
\tableline
Model I (J$^{\pi}$ E)$^{a)}$ & Model II (J$^{\pi}$ E)$^{b)}$ & Model III (J$^{\pi}$ E)$^{c)}$ & Model IV (J$^{\pi}$ E)$^{d)}$ \\
\tableline
\tableline
0$^{+}$ \hspace{1cm}0.000 &  0$^{+}$ \hspace{1cm}0.000 & 0$^{+}$ \hspace{1cm}0.000 &  0$^{+}$ \hspace{1cm}0.000  \\
2$^{+}$ \hspace{1cm}1.163 &  2$^{+}$ \hspace{1cm}1.015 & 2$^{+}$ \hspace{1cm}1.303 &  2$^{+}$ \hspace{1cm}1.253  \\
4$^{+}$ \hspace{1cm}2.790 &  4$^{+}$ \hspace{1cm}2.628 & 4$^{+}$ \hspace{1cm}2.741 &  4$^{+}$ \hspace{1cm}2.800  \\
6$^{+}$ \hspace{1cm}4.062 &  6$^{+}$ \hspace{1cm}4.079 & 6$^{+}$ \hspace{1cm}3.500 &  6$^{+}$ \hspace{1cm}3.738  \\
3$^{+}$ \hspace{1cm}5.786 &  8$^{+}$ \hspace{1cm}5.772 & 3$^{+}$ \hspace{1cm}4.716 &  3$^{+}$ \hspace{1cm}5.031  \\
5$^{+}$ \hspace{1cm}5.871 &  7$^{+}$ \hspace{1cm}6.018 & 5$^{+}$ \hspace{1cm}4.998 &  5$^{+}$ \hspace{1cm}5.082  \\
7$^{+}$ \hspace{1cm}6.043 & 12$^{+}$ \hspace{1cm}6.514 & 7$^{+}$ \hspace{1cm}5.356 &  7$^{+}$ \hspace{1cm}5.687  \\
8$^{+}$ \hspace{1cm}6.084 &  3$^{+}$ \hspace{1cm}6.540 & 8$^{+}$ \hspace{1cm}5.656 &  8$^{+}$ \hspace{1cm}6.045  \\
10$^{+}$\hspace{1cm}7.384 &  5$^{+}$ \hspace{1cm}6.602 & 9$^{+}$ \hspace{1cm}7.200 &  9$^{+}$ \hspace{1cm}7.731  \\
12$^{+}$\hspace{1cm}7.702 & 10$^{+}$ \hspace{1cm}6.722 &10$^{+}$ \hspace{1cm}7.200 & 10$^{+}$ \hspace{1cm}7.731  \\
 9$^{+}$\hspace{1cm}7.984 &  9$^{+}$ \hspace{1cm}8.048 &12$^{+}$ \hspace{1cm}7.840 & 12$^{+}$ \hspace{1cm}8.371  \\
\tableline
\end{tabular}
a) Input is spectrum of $^{42}$Sc(particle-particle)\\
b) Input is spectrum of $^{54}$Co(hole-hole)\\
c) The T=1 matrix elements are from the spectrum of $^{42}$Sc.  The T=0 matrix elements are degenerate at 1.5863 MeV.\\
d) Same as model 3 except that the J=1$^{+}$T=0 energy is lowered to 0.5863 MeV.\\
\end{table}

\newpage
\centerline{References}
\bigskip

\begin{enumerate}
\item A. Csoto and R.G. Lovas, Phys. Rev \underline{C53}, (1996) 1444

\item D.C. Zheng, J.P. Vory and B.R. Barrett, Phys. Rev \underline{C53} (1996) 1447

\item D.R. Tilley et. al., Energy Levels of Light Nuclei A=6, preliminary version 2

\item J.L. Forest et. al. Phys Rev \underline{C54}, (1996) 646

\item L. Zamick, D.C. Zheng and M. Fayache, Phys Rev \underline{C51},(1995) 1251

\item D.C. Zheng, B.R. Barrett, J.P. Vary and H. Muther Phys Rev \underline{C51}, (1995) 2471

\item S. Karataglidis, B.A. Brown, K. Amos and P.J. Dortmans, Phys. Rev \underline{C55} (1997) 2826

\item D.C. Zheng, B.R. Barrett, J.P. Vary, W.C. Haxton and C.-L. Song, Phys. Rev \underline{C52},2488 (1995)

\item M.S. Fayache, L. Zamick, and Y.Y. Sharon, Phys. Rev \underline{C55} (1997) 2389

\item M.S. Fayache, E. Moya de Guerra, P Sarrigueren, Y.Y. Sharon, and L. Zamick, Phys Rev \underline{C57}, (1998) 2351

\item F.C. Barker, Aust. J. Phys \underline{42}, (1989)25; \underline{41}(1988)743

\item P. Hodgeson, Survey Talk, Atomic and Nuclear Clusters Proceedings of the 
Second International Conference at Santorini, Greece June 28th - July 2 1993 Z. Phys A \underline{349},(1994) 197

\item D.F. Geesaman, R. Malmin, R.L. McGrath and J.W. Noe, P.R.L. \underline{34}, (1975) 326

\item D.F. Geesaman, R.L. McGrath, J.W. Noe, and R.E. Malmin, Physc. Rev \underline{19}, (1979) 1938

\item C.A. Ur et. al. Phys. Rev C \underline{58},(1998) 3163

\item D.C. Zheng, D. Berdichovsky, and L. Zamick, Phys Rev  \underline{C38} (1988) 437; 
J. Phys Soz Japan \underline{158} (1989) Suppl. 649; Phys Rev  \underline{C42} (1990) 1408

\item Y. Durga Devi, S. Robinson, and L. Zamick, Phys Rev \underline{C61} (2000) 037305

\item M. Hasegawa and K. Kaneko, Phys Rev  \underline{C61} (2000) 037306

\item L. Zamick and Y. Durga Devi, Phys Rev  \underline{C60}, (1999) 054317

\item J.D. McCullen, B.F. Bayman and L. Zamick, Phys. 
Rev. \underline{134B}, 515 (1964); Technical Report \underline{NYO-9891}.

\item R. Ernst, K.-H. Speidel, O. Kenn, U. Nachman, J. Gerber, P. Maier-Komor, N. Benczer-Koller, G. Jacob, G. Kumbartzki, L. Zamick, and F. Nowacki
 , Phys. Rev. Lett. \underline{84}, (2000) 416

\end{enumerate}

\end{document}